\documentclass{article}
\usepackage{dcase2024,amsmath,amssymb,graphicx,url,times,booktabs, tabularx,bm}

\title{Leveraging Self-supervised Audio Representations for Data-Efficient Acoustic Scene Classification}

\name{Yiqiang Cai$^{1}$,
      Shengchen Li$^{1}$, 
      Xi Shao$^{2}$
      }
\address{$^1$ Xi'an Jiaotong-Liverpool University, School of Advanced Technology, Suzhou, China,\\ yiqiang.cai21@student.xjtlu.edu.cn, shengchen.li@xjtlu.edu.cn\\          
        $^2$ Nanjing University of Posts and Telecommunications,\\ College of Telecommunications and Information Engineering, Nanjing, China, \\
        shaoxi@njupt.edu.cn\\
 }

\begin{document}

\ninept
\maketitle

\begin{sloppy}

\begin{abstract}
Acoustic scene classification (ASC) predominantly relies on supervised approaches. However, acquiring labeled data for training ASC models is often costly and time-consuming. Recently, self-supervised learning (SSL) has emerged as a powerful method for extracting features from unlabeled audio data, benefiting many downstream audio tasks. This paper proposes a data-efficient and low-complexity ASC system by leveraging self-supervised audio representations extracted from general-purpose audio datasets. We introduce BEATs, an audio SSL pre-trained model, to extract the general representations from AudioSet. Through extensive experiments, it has been demonstrated that the self-supervised audio representations can help to achieve high ASC accuracy with limited labeled fine-tuning data. Furthermore, we find that ensembling the SSL models fine-tuned with different strategies contributes to a further performance improvement. To meet low-complexity requirements, we use knowledge distillation to transfer the self-supervised knowledge from large teacher models to an efficient student model. The experimental results suggest that the self-supervised teachers effectively improve the classification accuracy of the student model. Our best-performing system obtains an average accuracy of 56.7\%\footnote{\url{https://github.com/yqcai888/easy_dcase_task1}}.
\end{abstract}


\section{Introduction}
\label{sec:intro}

Acoustic Scene Classification (ASC) is a task to recognize the environment in which an audio recording was captured, such as streets, parks, or airports \cite{barchiesi2015acoustic}. Traditional approaches to ASC typically rely on supervised learning techniques \cite{Kim2021b, Schmid2023, Cai2023a, tan2024acoustic}, which require large, labeled datasets to perform effectively. However, obtaining such labeled datasets is a resource-intensive process, often involving extensive manual annotation and data collection efforts. In the task 1 of the DCASE 2024 Challenge, participants are required to create low-complexity ASC systems that are trained with limited labeled data \cite{schmid2024data}. Specifically, five training subsets are provided, including 5\%, 10\%, 25\%, 50\%, and 100\% of the original training set's size. The performance of the submitted systems, trained on 5 subsets, is assessed by the average accuracy. This task encourages the development of efficient models capable of maintaining high performance despite reduced training data, advancing the practical applicability and scalability of ASC systems in real world.

In recent years, self-supervised learning (SSL) has been widely applied to address the scarcity of labeled data in audio tasks. SSL leverages the structure of the data to create supervisory signals, allowing models to learn meaningful representations from unlabeled audio data. SSAST \cite{Gong2022} introduces a masking strategy on the input spectrogram patches, allowing the transformer model to be pre-trained using both reconstruction loss and contrastive loss. Similarly, Audio-MAE \cite{Huang2022} and MaskSpec \cite{chong2023masked} pre-train an encoder-decoder transformer architecture by reconstructing the original audio spectrogram from its masked version. BEATs \cite{pmlr-v202-chen23ag} focuses on pre-training the transformer encoder by predicting the discrete labels generated by an acoustic tokenizer. After SSL pre-training on the general-purpose datasets, these models can be fine-tuned for various labeled tasks, such as keyword spotting and sound event detection. However, the application of audio self-supervised pre-trained models to ASC has been relatively unexplored.

In this work, we propose a data-efficient and low-complexity system with audio self-supervised pre-trained models for ASC. In Section \ref{sec:ssl}, BEATs \cite{pmlr-v202-chen23ag}, an audio transformer model SSL pre-trained on AudioSet \cite{gemmeke2017audio}, is introduced. The pre-trained encoders of models are then appended with a new linear classifier and fine-tuned on the ASC dataset. We experiment with various fine-tuning strategies and data augmentation techniques. The results demonstrate that the self-supervised representations extracted from the general-purpose audio dataset can significantly improve the ASC accuracy with limited labeled data. Moreover, it has been found that the ensemble of SSL models fine-tuned with different strategies makes further improvements to the ASC performance. Section \ref{sec:low-com} focuses on addressing the complexity requirements, where a knowledge distillation framework \cite{Schmid2023} is used to transfer the self-supervised knowledge from BEATs to TF-SepNet-64 \cite{cai2024tf, Cai2024}, which is an efficient CNN-based ASC model. The experimental results and ablation study are detailed in Section \ref{sec:result}. It shows that the self-supervised teachers significantly improve the performance of student model, achieving an average accuracy of 56.7\%. Our submitted system ranked 4th in the DCASE 2024 Challenge  \cite{Cai2024}.

\section{Self-supervised Pre-training and Fine-tuning}
\label{sec:ssl}
In this section, we aim to achieve high ASC accuracy with limited labeled data by leveraging the self-supervised audio representations. Specifically, we introduce BEATs, a state-of-the-art audio SSL model, to extract the general features from AudioSet \cite{gemmeke2017audio}. The SSL pre-trained models are then experimented with two fine-tuning strategies, frozen fine-tuning and unfrozen fine-tuning, for adapting to the ASC task. Experimental results are presented in Section \ref{ssec:accbeats}.

\begin{figure}[t]
\centering
\includegraphics[width=\linewidth]{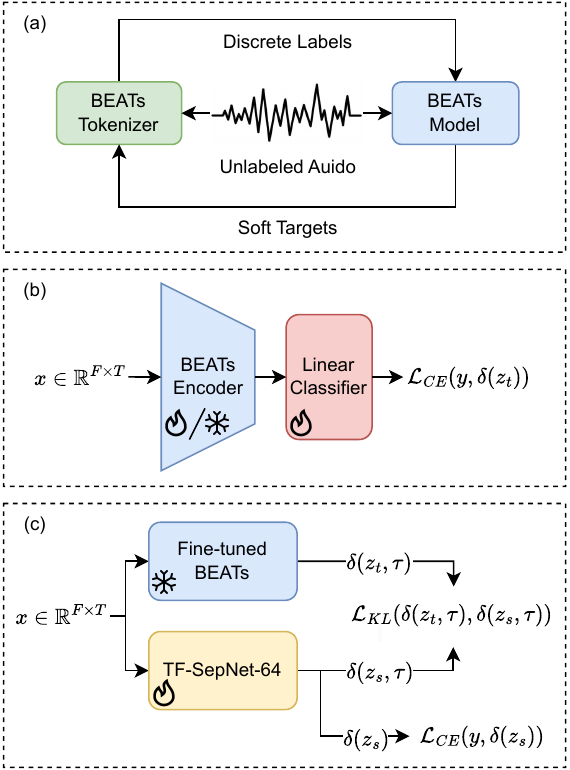} 
\caption{Proposed data-efficient and low-complexity ASC system. (a) Self-supervised pre-training BEATs on AudioSet. (b) Fine-tuning pre-trained BEATs on ASC dataset. (c) Distilling knowledge from fine-tuned BEATs to TF-SepNet-64. \textbf{Snowflake} icon indicates that the parameters of the corresponding part are frozen, while \textbf{Flame} icon indicates the opposite.}
\label{fig:system}
\end{figure}

\subsection{BEATs}
\label{ssec:beats}
Bidirectional Encoder representation from Audio Transformers (BEATs) \cite{pmlr-v202-chen23ag} is an audio pre-training framework that iteratively optimizes an acoustic tokenizer and an audio SSL model. As illustrated in Figure \ref{fig:system} (a), the BEATs tokenizer generates discrete labels of unlabeled audio, which the BEATs model learns to predict. Concurrently, the tokenizer is trained by distilling knowledge from the pre-trained BEATs model, enabling iterative optimization of both components. The authors argue that discrete label prediction captures high-level audio semantics  more effectively than the reconstruction loss used in previous audio SSL models. The tokenizer and label predictor of SSL model are discarded after self-supervised pre-training. In the original work, BEATs models are self-supervised pre-trained, and alternatively supervised fine-tuned, on AudioSet before applying to downstream tasks. For convinience, we denote the purely self-supervised pre-trained BEATs model as BEATs (SSL). The SSL pre-trained BEATs with additional supervised fine-tuning on AudioSet is denoted as BEATs (SSL+SL).

Before fine-tuning for ASC, the reserved BEATs encoder is appended with a task-specific linear classifier to output class probabilities for different acoustic scenes, as shown in Figure \ref{fig:system} (b). The linear classifier consists of a linear layer, a mean-pooling layer and a softmax operation. The fine-tuning data is from TAU Urban Acoustic Scene 2022 Mobile development dataset \cite{Heittola2020}. Each audio clip is resampled to 16 kHz, and 128-dimensional Mel-filter bank features are extracted using a 25 ms Povey window with a 10 ms shift. The features are normalized according to the mean and standard deviation of AudioSet. Each acoustic feature $x\in \mathbb{R}^{F\times T}$ is then divided into 16 × 16 patches and flattened into a sequence of patches to serve as input for the pre-trained BEATs model.

\begin{table}[t]
    \centering
    \begin{tabular}{l|c c c c c|c}
    \toprule
    \textit{Model}& 5\%& 10\%& 25\%& 50\%& 100\%& Avg.\\
    \midrule
    BEATs (SSL)\text{*}& 50.7& 52.0& 54.2& 55.0& 55.8& 53.5\\
    BEATs (SSL)& 52.9& 54.9& 58.1& 59.7& 61.2& 57.4\\
    BEATs (SSL+SL)& 54.3& 56.6& 59.7& 60.7& 62.1& 58.7\\
    \midrule
    3 Ensemble& 55.4& 57.6& 61.1& 62.2& 64.2& 60.1\\
    12 Ensemble& \textbf{55.8}& \textbf{58.0}& \textbf{61.6}& \textbf{62.9}& \textbf{64.6}& \textbf{60.6}\\
    \bottomrule
    \end{tabular}
    \caption{Accuracy of fine-tuned BEATs on the test set of TAU Urban Acoustic Scene 2022 Mobile development dataset \cite{Heittola2020}. \textbf{SSL} denotes the BEATs model is self-supervised pre-trained on AudioSet. \textbf{SSL+SL} denotes the SSL pre-trained BEATs model is additionally supervised fine-tuned on AudioSet. \textbf{*} indicates the encoder of BEATs is frozen during the fine-tuning on ASC dataset. Top-1 accuracy of 5 independent runs is presented.}
    \label{tab:accbeats}
\end{table}

\subsection{Frozen Fine-tuning}
\label{ssec:frozen}
To evaluate the benefits of self-supervised audio representations, the encoder of BEATs (SSL) is frozen as a feature extractor while only the linear classifier is trained with the cross entropy loss, as shown in Figure \ref{fig:system} (b). The frozen model is denoted as BEATs (SSL)*. Frozen fine-tuning allows the model to leverage representations learned during self-supervised pre-training, preventing overfitting and catastrophic forgetting \cite{davari2022probing}. We train BEATs (SSL)* for 60 epochs using the default Adam optimizer. To further enhance the robustness and generalization of the model, we apply two widely-used data augmentation methods: Mixup \cite{zhang2018mixup} with an $\alpha$ of 0.3 and SpecAugmentation \cite{Park2019} with a mask ratio of 0.2.

\subsection{Unfrozen Fine-tuning}
Beside freezing the SSL models as feature extractors, we also explore unfrozen fine-tuning to further adapt BEATs to the ASC task. Unfrozen fine-tuning allows the model to refine representations learned during self-supervised pre-training, typically leading to better performance  compared to frozen fine-tuning.

We apply BEATs (SSL) and BEATs (SSL+SL) for unfrozen fine-tuning, using the same training configurations. The models are fine-tuned for 30 epochs with a batch size of 512. The AdamW optimizer \cite{loshchilov2018decoupled} is applied with $\beta$ = (0.9, 0.98) and a weight decay of 0.01. The learning rate is scheduled to exponentially increase from 0 to a peak value of $1\times10^{-5}$ over four epochs, then linearly decrease to a minimum value of $5\times10^{-8}$ for the remaining epochs. Four data augmentation techniques are used during fine-tuning: Mixup \cite{zhang2018mixup} with $\alpha$ = 0.3, Freq-MixStyle \cite{Schmid2022} with $\alpha$ = 0.4 and $p_{fms}$ = 0.4, SpecAugmentation \cite{Park2019} with a mask ratio of 0.2, and DIR augmentation \cite{morocutti2023device} with $p_{dir}$ = 0.6.

\subsection{Ensemble Models}
Previous works \cite{Schmid2023,Schmid2022} have shown that model ensemble with different configurations can enhance ASC performance and benefit knowledge distillation. In this work, we average the logits to ensemble BEATs models that fine-tuned with different fine-tuning strategies. The small ensemble consists of three fine-tuned BEATs models: BEATs (SSL)*, BEATs (SSL) and BEATs (SSL+SL). The large ensemble includes twelve fine-tuned BEATs models: one BEATs (SSL)*, one BEATs (SSL) and ten BEATs (SSL+SL). The ten BEATs (SSL+SL) models are AudioSet fine-tuned BEATs models with different tokenizers as described in the original work \cite{pmlr-v202-chen23ag}.

\section{Knowledge Distillation with Self-supervised Teachers}
\label{sec:low-com}

DCASE Challenge 2024 task 1 imposes strict limitations on computational complexity, restraining the model size within 128kB and the number of multiply-accumulate operations within 30 MMACs. In this section, knowledge distillation \cite{gou2021knowledge} is introduced to transfer knowledge from the fine-tuned BEATs to an efficient student model, TF-SepNet-64. By employing the self-supervised teachers, we aim to develop ASC systems that operate within the computational limits while maintaining high accuracy with limited labeled data. The framework of proposed system is shown in Figure \ref{fig:system} (c).

\subsection{TF-SepNet-64}
\label{ssec:student}
Time-Frequency Separate Network (TF-SepNet) \cite{cai2024tf} is a deep CNN architecture designed specifically for low-complexity ASC tasks, achieving second place in DCASE Challenge 2023. TF-SepNet processes features separately along the time and frequency dimensions using one-dimensional (1D) kernels, which reduce computational costs and provide a larger effective receptive field (ERF), allowing the model to capture more time-frequency features.

As in \cite{Cai2024}, TF-SepNet-64 is optimized to meet the upper complexity limit of the challenge requirements. Several adjustments have been made. First, the number of base channels is set to 64. Second, all Adaptive Residual Normalization layers \cite{Cai2023a} are replaced with Residual Normalization layers \cite{Kim2021b} to reduce the number of model parameters. Third, a Max-pooling layer is added before the last TF-SepConvs block to further reduce the feature size. In the finish, the total parameter number of TF-SepNet-64 is 126,858. For an input feature size of (512, 64), the maximum number of MACs per inference is 29.4196 MMACs.

\subsection{Knowledge Distillation}
\label{ssec:kd}

We adopt the widely used knowledge distillation framework in previous years' challenges \cite{Schmid2023,Schmid2022}, which focuses on directly mimicking the final predictions of the teacher model. As illustrated in Figure \ref{fig:system} (c), the knowledge transfer involves two main steps.

The input feature is a log-mel spectrogram $x\in \mathbb{R}^{F\times T}$. For the teacher path, once the self-supervised teachers are fine-tuned, as shown in Figure \ref{fig:system} (b), the predictions on a specified training subset are computed, serving as the teacher logits in the knowledge distillation process. For the student path, the ASC student is trained on the specified training subset using a combination of the ground truth labels and the soft targets provided by the teacher model. Give a vector of logits $z$ as the outputs of the last classification layer of a model, the soft targets are the probabilities that the input belongs to the classes and can be estimated by a softmax function $\delta(\cdot)$ as

\begin{equation}
    \delta(z_i, \tau)=\frac{\exp(z_i/\tau)}{\sum_j \exp(z_j/\tau)}
\end{equation}

\noindent
where $z_i$ is the logit for the i-th class, and a temperature factor $\tau$ is introduced to control the importance of each soft target. The training objective of student model is to minimize the divergence between the student’s predictions and the soft targets from the teacher, as well as to correctly classify the labeled data. The overall loss function for the student can be formulated as

\begin{equation}
    \mathcal{L}=\lambda \mathcal{L}_{CE}(y, \delta(z_s))+(1-\lambda)\tau^2\mathcal{L}_{KL}(\delta(z_t, \tau), \delta(z_s, \tau))
\end{equation}

\noindent
where $\mathcal{L}_{CE}$ is the cross-entropy loss between the ground truth labels and the student’s predictions, and $\mathcal{L}_{KL}$ is the Kullback-Leibler divergence between the soft targets from the teacher and the student’s predictions. $\lambda$ is a hyperparameter to balance the weight between label and distillation loss.

\subsection{Experimental Setup}
\label{sec:setup}

\noindent
\textbf{Dataset and Baseline:} The dataset for the task1 of DCASE 2024 Challenge has exactly the same content as the TAU Urban Acoustic Scenes 2022 Mobile development dataset \cite{Heittola2020}, but the training sets of different sizes are provided. These train subsets contain approximately 5\%, 10\%, 25\%, 50\%, and 100\% of the audio snippets in the training set provided in previous years. The DCASE baseline model for comparison, CP-Mobile \cite{Schmid2023workshop}, is a fully-supervised CNN classifier that achieved top ranking in DCASE Challenge 2023.

\noindent
\textbf{Feature Extraction:} For TF-SepNet-64, we generally follow the baseline settings \cite{Schmid2023workshop} for feature extraction. The audio recordings are firstly resampled to 32 kHz. Time-frequency representations are then extracted using a 4096-point FFT with a window size of 96 ms and a hop size of 16 ms. The primary difference in our approach is the application of a Mel-scaled filter bank with a large number of frequency bins, 512, to convert the spectrograms into mel spectrograms, which leads to a slight improvement on the classification accuracy. The final input size for TF-SepNet-64 is (512, 64).

\noindent
\textbf{Data Augmentations:} Data augmentation is a crucial technique in ASC tasks, especially when the labeled data is limited. In this work, we use a combination of Soft Mixup \cite{Cai2024}, Freq-MixStyle \cite{Schmid2022}, and Device Impulse Response (DIR) augmentation \cite{morocutti2023device} to enhance the diversity and quality of our training data. $\alpha$ of Soft Mixup is set to 0.3. $\alpha$ and $p$ of Freq-MixStyle are respectively set to 0.4 and 0.8. $p_{dir}$ of DIR augmentation is set to 0.4. All augmentations are implemented to be plug-and-played during training.

\noindent
\textbf{Training:} We train TF-SepNet-64 for 150 epoch using Adam optimizer with different initial learning rate for 5 subsets, 0.06 for split5, 0.05 for split50 and 0.04 for all other splits. Stochastic Gradient Descent with Warm Restarts (SGDR) \cite{loshchilov2017sgdr} is applied with $T_0$ =10 and $T_{mult}$ = 2, where the learning rate is periodically reset to initial value and then decayed with cosine annealing. The batch size is set to 512.  We fix $\lambda=0.02$ and $\tau=2$ for the knowledge distillation as in \cite{Schmid2023}. After training, Post-Training Static Quantization is implemented through the Intel Neural Compressor\footnote{https://intel.github.io/neural-compressor} to quantize the weights of model into INT8 data type.

\begin{table}[t]
    \centering
    \begin{tabular}{l|c c c c c|c}
    \toprule
    \textit{Model}& 5\%& 10\%& 25\%& 50\%& 100\%& Avg.\\
    \midrule
    DCASE Baseline& 42.4& 45.3& 50.3& 53.2& 57.0& 49.6\\
    \midrule
    TF-SepNet-64& 45.7& 51.1& 55.6& 59.6& 62.5& 54.9\\
    +BEATs (SSL)\text{*}& 48.2& 51.0& 54.9& 58.0& 59.9& 54.4\\
    +BEATs (SSL)& 47.3& \textbf{52.5}& 57.6& 60.8& 61.9& 56.0\\
    +BEATs (SSL+SL)& 47.8& 52.1& 57.7& \textbf{61.1}& 62.6& 56.3\\
    +3 Ensemble& \textbf{49.0}& 52.3& \textbf{57.9}& 60.7& \textbf{63.5}& \textbf{56.7}\\
    +12 Ensemble& 47.9& 52.3& 57.5& 60.1& 62.8& 56.1\\
    \bottomrule
    \end{tabular}
    \caption{Accuracy of TF-SepNet-64 with different BEATs teachers on the test set of TAU Urban Acoustic Scene 2022 Mobile development dataset \cite{Heittola2020}. The teacher logits of each BEATs model is used (\textbf{+}) in knowledge distillation at a time. Top-1 and quantized accuracy of 5 independent runs is presented.}
    \label{tab:acctf}
\end{table}

\begin{figure}
    \centering
    \includegraphics[width=\linewidth]{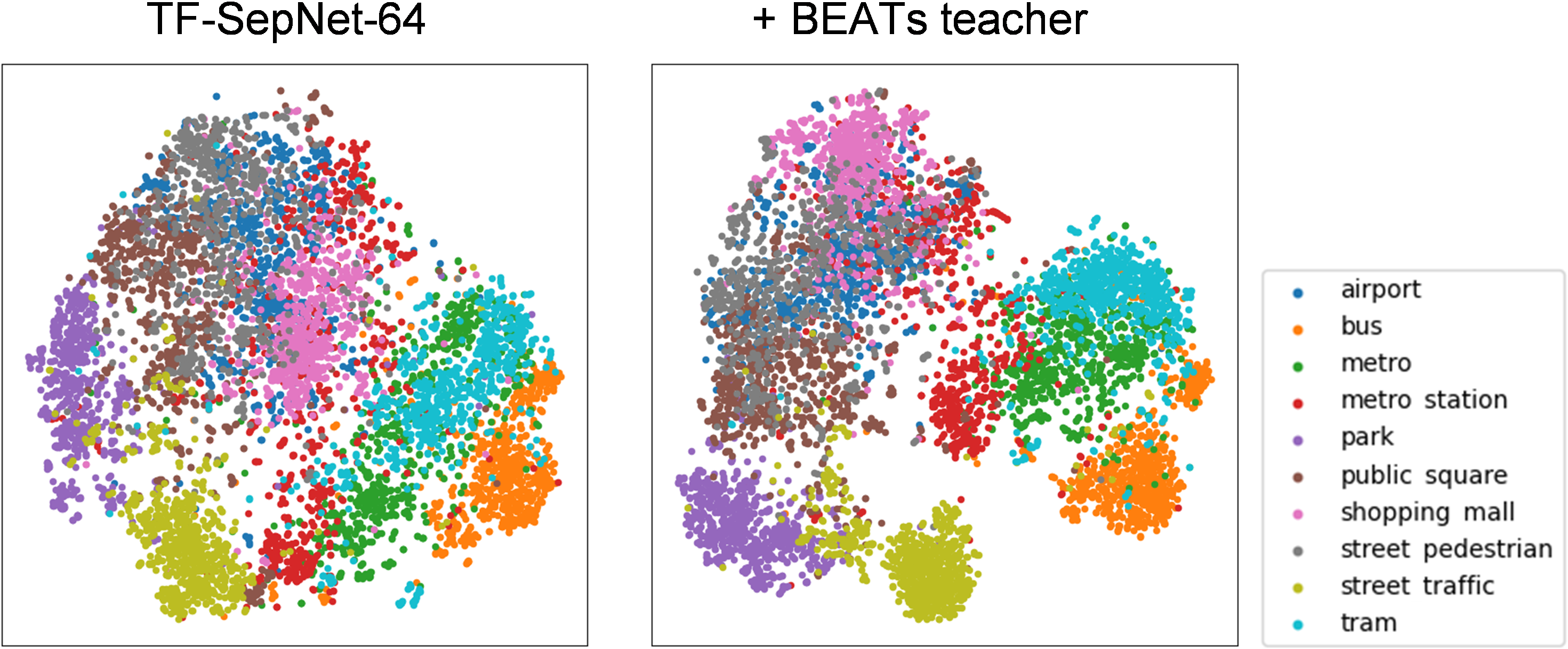}
    \caption{TSNE \cite{van2008visualizing} visualization of acoustic scene features extracted by TF-SepNet-64, which is trained on the 5\% subset. \textbf{Left:} Knowledge distillation is not applied. \textbf{Right:} Distilling knowledge from the 3 ensemble BEATs teacher.}
    \label{fig:tsne}
\end{figure}

\section{Results}
\label{sec:result}

\subsection{Performance of Fine-tuned BEATs}
\label{ssec:accbeats}
Table \ref{tab:accbeats} presents the accuracy of fine-tuned BEATs using different fine-tuning strategies. Even with the encoder frozen, BEATs (SSL)* achieves over 50\% accuracy with only 5\% training data. This result demonstrates the self-supervised representations learned from general-purpose audio dataset are beneficial to the ASC task, especially when labeled data is exceptionally limited. However, the accuracy witnesses little improvements with the increase of training data. This is due to the limited capability of a single linear layer to adapt to changes in data scale. When the encoder is unfrozen during fine-tuning, BEATs (SSL) shows a significant 3.9\% improvement in average accuracy. Additionally, the AudioSet supervised fine-tuned model, BEATs (SSL+SL), achieves further improvements. For the model ensembles, the 3 ensemble outperforms the best single model by 1.4\% in average accuracy, and the large 12 ensemble achieves an average accuracy of 60.6\%. The different fine-tuning strategies diversify the predictions for ensembling, effectively combining self-supervised knowledge and supervised knowledge.

\subsection{TF-SepNet-64 with BEATs Teachers}

The performance of TF-SepNet-64 with various BEATs teachers is shown in Table \ref{tab:acctf}. TF-SepNet-64 without knowledge distillation outperforms the DCASE baseline by 5.3\% in average accuracy but experiences considerable drop as the amount of training data decreases. The single BEATs (SSL)* teacher only helps in the 5\% subset while BEATs (SSL) and BEATs (SSL+SL) improve the student model across more subsets. By comparing the performance between TF-SepNet-64 and BEATs, we infer that a teacher model is generally helpful when it has a higher accuracy than the student. Nevertheless, BEATs (SSL) helps to obtain the highest accuracy in the 10\% subset while BEATs (SSL+SL) is most effective in the 50\% subset. Compared to individual teachers, the ensemble teachers generally provide greater benefits to the student. Interestingly, rather than the large 12 ensemble, the small 3 ensemble achieves the best performance for the remaining subsets, obtaining the highest average accuracy of 56.7\%. Therefore, a teacher with higher accuracy does not necessarily guarantee better improvement for the student. To further examine the benefits of BEATs teacher, we visualize the acoustic scene features as shown in Figure \ref{fig:tsne}. The samples are better clustered with the assistance of BEATs teacher.

\subsection{Ablation Study}
Table \ref{tab:ablation} presents the ablation study for our proposed system (TF-SepNet-64 + 3 BEATs ensemble) on the two extreme subset: 5\% and 100\%. The configurations for TF-SepNet-64, such as using a larger amounts of Mel bins, more base channels, replacing AdaResNorm with ResNorm, and adding a Max-pooling layer, contributes to performance improvements to varying degrees while maintaining the system's complexity within the challenge requirements. Meanwhile, the data augmentation methods enhance the accuracy without introducing additional overheads. The results also indicate that the BEATs teacher is the dominant factor in performance when labeled training data is extremely limited.

\begin{table}[t]
    \centering
    \begin{tabular}{l|c c|c c}
    \toprule
    \textit{System}& 5\%& 100\%& MMACs& Param/k\\
    \midrule
    \textbf{Proposed System}& \textbf{49.0}& \textbf{63.5}& 29.4& 126.9\\
    \midrule
    Mel bins (512$\rightarrow$256)& 47.3& 61.9& 14.8& 126.9\\
    Base channels (64$\rightarrow$40)& 45.8& 61.3& 12.9& 52.3\\
    ResNorm$\rightarrow$AdaResNorm& 48.8& 61.9& 29.4& 128.6\\
    w/o added Max-pooling& 45.9& 63.3& 32.0& 126.9\\
    \midrule
    w/o Soft Mixup& 46.0& 62.3& 29.4& 126.9\\
    w/o Freq-MixStyle& 47.0& 61.6& 29.4& 126.9\\
    w/o DIR Augmentaion& 48.0& 62.4& 29.4& 126.9\\
    \midrule
    w/o BEATs teacher& 45.7& 62.5& 29.4& 126.9\\
    \bottomrule
    \end{tabular}
    \caption{Ablation study of our proposed system (TF-SepNet-64 + 3 BEATs ensemble). Each component is changed ($\bm{\rightarrow}$) or removed (\textbf{w/o}) at a time. \textbf{MMACs} (million multiply-accumulate operations) represents the computational costs per inference. \textbf{Param/k} denotes the number of parameters.}
    \label{tab:ablation}
\end{table}

\section{Conclusion}
\label{sec:conclusion}

In this paper, we introduce self-supervised audio representations to address the challenge of data-efficient low-complexity acoustic scene classification (ASC). We fine-tune BEATs models as self-supervised teachers and then transfer the knowledge to a low-complexity student model, TF-SepNet-64, through a knowledge distillation framework. The experimental results demonstrate the effectiveness of self-supervised pre-trained models in the ASC task, and also show the benefits of self-supervised teachers for the low-complexity student model when the labeled training data is limited.

\section{Acknowledgement}
\label{sec:ack}

This project is supported by the Gusu Innovation and Entrepreneurship Leading Talents Programme (No: ZXL2022472).

\bibliographystyle{IEEEtran}
\bibliography{refs}

%
%
%
%
%
%
%
%
%

\end{sloppy}
\end{document}